\documentstyle[preprint,aps]{revtex}
\begin{document}
\draft
\input{epsf}

\title{To what extent can dynamical models describe statistical features of
turbulent flows?}

\author{V. Carbone$^1$, R. Cavazzana$^2$, V. Antoni$^{2,3}$,
L. Sorriso--Valvo$^1$, E. Spada$^2$, G. Regnoli$^{2,3}$, P. Giuliani$^5$,
N. Vianello$^{2,3}$, F. Lepreti$^1$, R. Bruno$^6$, E. Martines$^2$, P.
Veltri$^1$}

\address{$^1$Dipartimento di Fisica and Istituto di Fisica della Materia \\
Universit\`a della Calabria, 87036 Rende (CS), Italy. \\
$^2$Consorzio RFX - Associazione Euratom-ENEA per la fusione, Padova, Italy. \\
$^3$Istituto di Fisica della Materia, Unit\`a di Padova, Italy. \\
$^5$The Niels Bohr Institute and Danish Metereological Institute, DK--2100
Copenhagen, Denmark. \\
$^6$Istituto di Fisica dello Spazio Interplanetario,
Roma, Italy.}

\date{\today}
\maketitle

\begin{abstract}

Statistical features of "bursty" behaviour in charged and neutral fluid
turbulence, are compared to statistics of intermittent events in a GOY
shell model, and avalanches in different models of Self Organized Criticality
(SOC). It is found that inter--burst times show a power law distribution for
turbulent samples and for the shell model, a property which is shared only in a
particular case of the running sandpile model. The breakdown of
self--similarity generated by isolated events observed in the turbulent
samples, is well reproduced by the shell model, while it is absent in all SOC
models considered. On this base, we conclude that SOC models are not adequate
to mimic fluid turbulence, while the GOY shell model constitutes a better
candidate to describe the gross features of turbulence.

\end{abstract}

%\pacs{PACS Numbers: 05.45.-a; 05.65.+b; 52.25.Gj;}



\newpage
Turbulence in neutral and charged fluids exhibits bursty behaviour, power
laws and $1/f$ spectra. Understanding the origin of the rich and complex
dynamics underlying these properties, is a challenging and fascinating task.
In the years different models have been proposed aimed to catch the fundamental
mechanism of this phenomenology. Among them simplified models ("sandpiles")
based on Self--Organized Criticality (SOC) \cite{soc} and dynamical systems
approach \cite{Bohr}, are particularly suitable for this purpose. SOC has been
proposed as a paradigm for some complex systems, such as earthquakes, solar
flares, river floods, fusion plasmas, etc. Despite the very simple governing
rules, these systems  exhibit a rich dynamics showing bursty behaviour with
impulsive events ("avalanches" in sandpiles) which remind a variety of
natural phenomena, including turbulence. Power laws are observed for the total
energy \cite{soc,kadanoff,OFC}, the duration and the peak energy distributions
of the bursts \cite{soc} and for the total energy frequency spectra
\cite{1/f}. On the other hand models based on the dynamical systems approach
have been developed by reducing the fluid equations, in order to isolate the
basic mechanisms generating  turbulence. An  example of a model belonging to
this class is the so called GOY shell model (see \cite{Bohr} and ref.s
therein). This model describes
a one dimensional fluid system by a set of ordinary
differential equations of complex variables non linearly interacting through
the coupling of the neighbour and next-neighbour modes.

To what extent these simplified models are good candidates to mimic
turbulence is an intriguing question, reported, for example, in Ref.s
\cite{soc,noi,papero,lastprl}. The present paper is aimed to  contribute to
clarify this issue by  comparing the predictions of these different dynamical
models with samples of turbulence in neutral and charged fluids. The
comparison is carried out by discussing two basic properties of turbulence,
namely the presence of long time correlations and intermittency. For this
purpose two statistical analysis tools have been used: the Distribution of the
Waiting Times (WTD)  $\Delta t$ between subsequent bursts \cite{noi} and the
comparison of the Probability Distribution Function (pdf) of fluctuations
resolved at different time scales \cite{vanatta}. These methods have been
applied to SOC and GOY models as well as to experimental data. In particular,
concerning SOC models, we have analyzed succesive generalization of the
original sandpile  model (BTW model), made by Kadanoff  et al.
\cite{kadanoff} which varied the underlying microscopic rules and by Hwa and
Kardar \cite{running} which included finite drive effects (the running
sandpile).

It is worth noticing that the analysis of dynamical properties of these models
requires a correct definition of the timescale, a point often neglected in
SOC literature. For example Ref. \cite{kadanoff} does not give any definition
of the timescale, since it is focused on integral properties of avalanches,
namely the total released energy and particles. With this in mind we choose to
introduce in Kadanoff sandpile the same timescale definition proposed in Ref.
\cite{OFC}. Accordingly in this case we built the signal of dissipated energy
$\epsilon_{K}(t)$ considering every avalanche as an istantaneous event, with
an amplitude equal to the integral avalanche size and the time between two
subsequent events proportional to the number of particles needed to put the
system in the unstable state again. On the other hand in the case of the
running sanpile \cite{running} the dissipated energy signal and its timescale
are clearly defined. In each temporal step the system is continuously fed at
random with a certain finite deposition rate probability $J_{in}$ and the
unstable sites are simultaneously updated. The dissipated energy signal
$\epsilon_{H}(t)$ in this case is taken as the total number of unstable sites
at each step.  The timescale of the previous model is recovered in the limit
$J_{in} \to 0$. In our simulation we consider a sandpile of size $\ell$, and
associate to each lattice site $n$ the height of the pile $h(n,t)$. If there
is any unstable site according to the rule $h(n,t) - h(n \pm 1,t) > \Delta $,
the system is evolved by taking a fixed amount $N_f$ of sand from the unstable
$n$ site and redistributing grains to the neighbouring $n \pm 1$ site (we set
the threshold $\Delta = 50$ and $N_f= 20$).
%If feeding is suspended during
%unstable evolution of the system Kadanoff model is obtained, otherwise the
%running sand-pile model is recovered.

\begin{figure}
\epsfxsize=12cm
\centerline{\epsffile{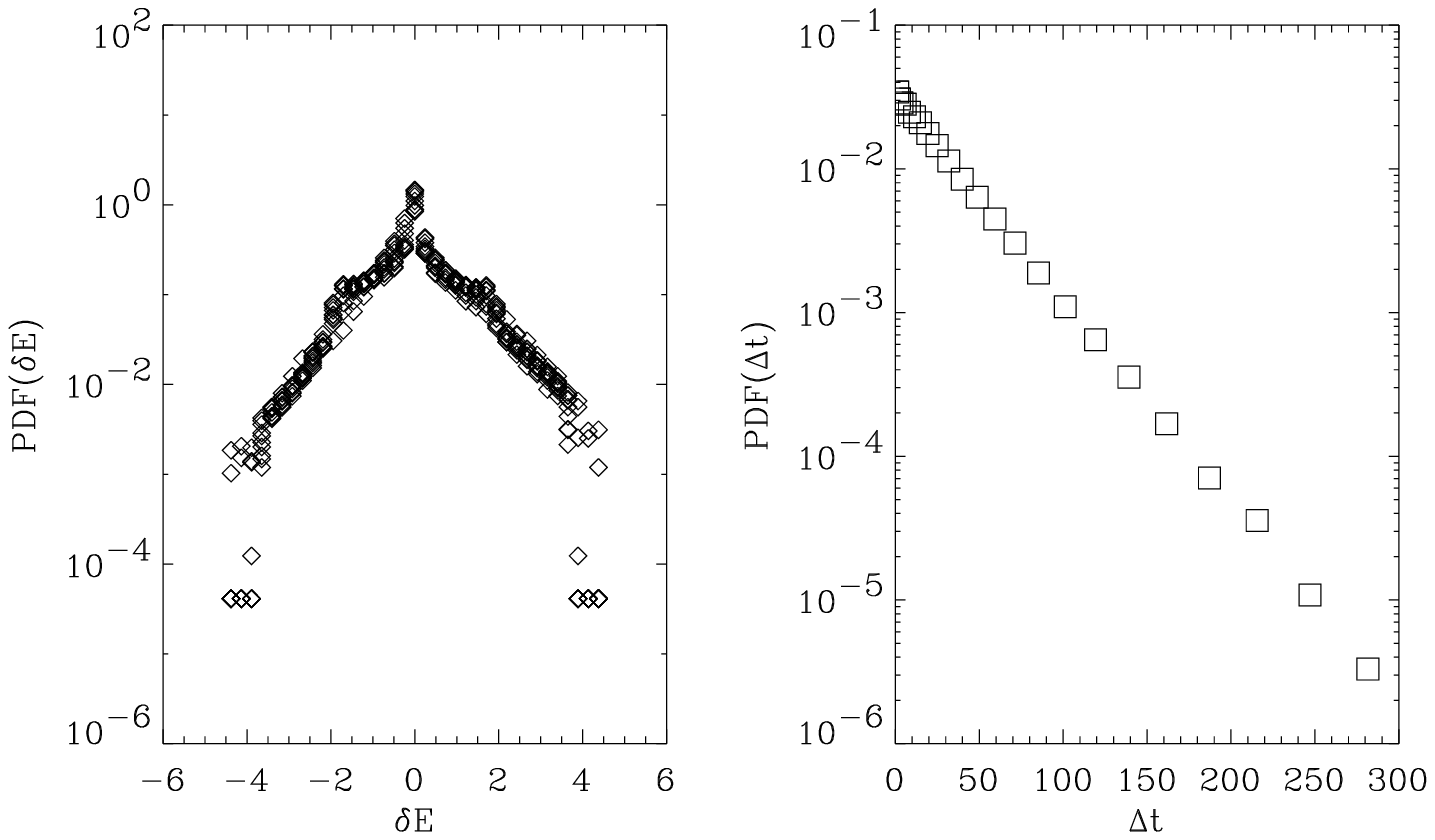}}
\caption{We show the results obtained from the Kadanoff ``local limited''
 sandpile obtained with a system size $\ell = 1024$. In the left
panel we show the pdfs of the rescaled fluctuations at different scales
$\tau$, and in the right panel (log--lin plot) we show the distribution of
waiting times between avalanches. Simulations obtained with different values
of $\ell$ show the same results.}
\label{syslength}
\end{figure}

The WTD $P(\Delta t)$ for the Kadanoff  model (right--hand panel of
Fig. \ref{syslength}) shows a clear exponential decay. This behaviour, which is
known to be associated to independent Poissonian events \cite{noi}, indicates
that no long term correlations are present between subsequent avalanches, as
found in the standard BTW model \cite{noi}.

\begin{figure}
\epsfxsize=12cm
\centerline{\epsffile{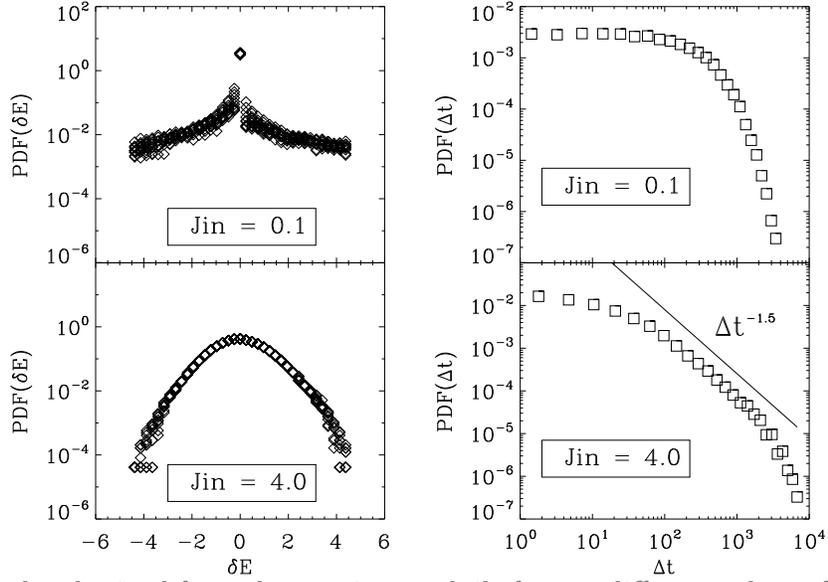}}
\caption{Results obtained from the running sandpile for two different
values of $J_{in}$ (see text). In the left panels we show the pdfs of the
rescaled fluctuations at different scales $\tau$, and in the right panels
(log--log plot) we show the distribution of waiting times between avalanches.}
\label{running}
\end{figure}

In the case of running sanpile at small $J_{in}$, an exponential WTD was
found (top right--hand panel of Fig. \ref{running}), recovering the behaviour
of the Kadanoff sanpile. On the other hand, increasing the drive $J_{in}$ a
tendency to a power law behaviour $P(\Delta t) \sim \Delta t^{-1.5}$ emerges.
In order to measure the waiting times, we note that while at $J_{in} \to 0$
avalanches are isolated events so that waiting times coincide with quiescent
intervals, at high $J_{in}$ there is a continuos activity. Thus to define the
waiting times between subsequent events a threshold has been chosen applying
the algorithm reported in \cite{noi}. The waiting time corresponds to the
time interval between two subsequent events with amplitude beyond this
threshold. In the limit $J_{in} \to 0$ this method correctly gives threshold
$0$.

It is known that long range time correlations can be obtained by overlapping
signals from uncorrelated avalanches \cite{Krommes,Jensen}. In order to test
the origin of the power law above reported, following a procedure originally
proposed in ref. \cite{soc}, a signal has been generated by randomly
overlapping isolated avalanches for the case $J_{in} \to 0$, keeping an
average dissipation rate equal to that  for the case $J_{in} = 4$. In this
case an exponential decay of the WTD has been found. Thus we conclude that
the origin of power law in the WTD of the Hwa--Kardar model is due to the
interaction among the avalanches inside the system. The emerging of a power
law behavior with exponent close to $\beta \simeq 1.5$ in the WTD is a common
feature found for all Hwa--Kardar systems, indipendently from their size
$\ell$.

In order to characterize the properties of the dynamics of these SOC systems,
the energy fluctuations $\delta \epsilon_{\tau} =
\epsilon(t+\tau)-\epsilon(t)$ at different scales $\tau$ of the energy
signal $\epsilon(t)$ defined above have been analysed. The increments $\delta
\epsilon_{\tau}$ of a stochastic process are self--similar with respect to the
scaling transformation $\tau \to \lambda \tau$ (with $\lambda > 0$) if $\delta
\epsilon_{\tau} \sim \lambda^{-\alpha} \delta \epsilon_{\lambda \tau}$
\cite{vanatta}, which means that the cumulative distributions of the first and
the second term coincide for a given scaling exponent $\alpha$. This can be
translated on the relationship between density probability of increments at
two different scales $\tau$, namely $pdf(\delta \epsilon_{\tau}) =
\lambda^{-\alpha} pdf(\lambda^{-\alpha} \delta \epsilon_{\lambda \tau})$. Test
for self--similarity of random processes can be made without actually assuming
a value for $\alpha$. In fact it can be shown that pdfs of rescaled
fluctuations $\delta E_{\tau} = \delta
\epsilon_{\tau}/\sigma_{\epsilon}(\tau)$, where $\sigma_{\epsilon}(\tau) = \{
\langle[\epsilon(t+\tau)-\epsilon(t)]^2\rangle \}^{1/2}$ (brackets being time
averages), are invariant in a self--similar process, say $pdf(\delta E_{\tau})
= pdf(\delta E_{\lambda \tau})$ \cite{vanatta}.

The pdfs of the fluctuations $\delta E_{\tau}$ at different scales $\tau$
for the  Kadanoff sandpile are shown on the left--hand panel of
Fig. \ref{syslength}. The bursty appereance of SOC signals is reflected onto
the shape with fat tails of the pdfs. However pdfs are invariant when plotted
in the rescaled form, so the process turns out to be purely self--similar.

Concerning the running sandpile, in the case of small drive ($J_{in} \to 0$)
the result turns out to be close (upper--left panel of Fig. \ref{running}) to
the Kadanoff model, confirming that in the above limit the two models
are equivalent.

In the case of finite drive (bottom--left panel of Fig. \ref{running}), the
shape of pdfs approaches a Gaussian shape due to the interaction of the
avalanches, and, even more noticeable,  the underlying process remains purely
self--similar.  These results are consistent with the intrinsic self--similar
structure of SOC models \cite{soc}, which is at the origin of its power law
distributions and scaling properties \cite{kadanoff}. It is worth  noticing
that self similarity is observed even in the case of finite drive, despite
the common belief  that '' the true SOC behaviour arises only in the limit
that the  forcing rate approaches 0'' \cite{Krommes,Jensen}.

The MHD GOY shell model \cite{giuliani} describes the dynamic of the energy
cascade in turbulence. The wavevector space is divided in $N$ shells, each
shell being characterised by a discrete wavevector $k_n = 2^n k_0$ ($n =
0,1,...,N$), and by the (complex) dynamical variables, velocity $u_n(t)$ and
magnetic field $b_n(t)$, which represent characteristic fluctuations across
eddies at the scale $\ell_n \sim k_n^{-1}$. The dynamical behaviour of the
model is described by the following set of ordinary differential equations

\begin{eqnarray}
{du_n \over dt} = -\nu k_{n}^{2}u_{n}+ik_{n}\Bigl\{
(u_{n+1}u_{n+2}-b_{n+1}b_{n+2}) \nonumber \\
- \frac{1}{4}(u_{n-1}u_{n+1}-b_{n-1}b_{n+1})
- \frac{1}{8}(u_{n-2}u_{n-1}-b_{n-2}b_{n-1})
\Bigr\}^{\ast}+f_{n}
\label{nonlinearev}
\end{eqnarray}
\begin{eqnarray}
{db_n \over dt} = -\eta k_{n}^{2}b_{n}+\frac{ik_n}{6}\Bigl\{
(u_{n+1}b_{n+2}-b_{n+1}u_{n+2})\nonumber \\
+(u_{n-1}b_{n+1}-b_{n-1}u_{n+1})+
(u_{n-2}b_{n-1}-b_{n-2}u_{n-1})
\Bigr\}^{\ast}
\label{nonlineareb}
\end{eqnarray}
$\nu$ and $\eta$ are the kinematic viscosity and the resistivity
respectively, and $f_n$ is an external forcing term. The nonlinear terms have
been obtained by imposing quadratic nonlinear coupling between neighbouring
shells and the conservation of three ideal invariants. Detailed informations
on the model and on simulations can be found in Ref.s \cite{noi,giuliani}.

\begin{figure}
\epsfxsize=12cm
\centerline{\epsffile{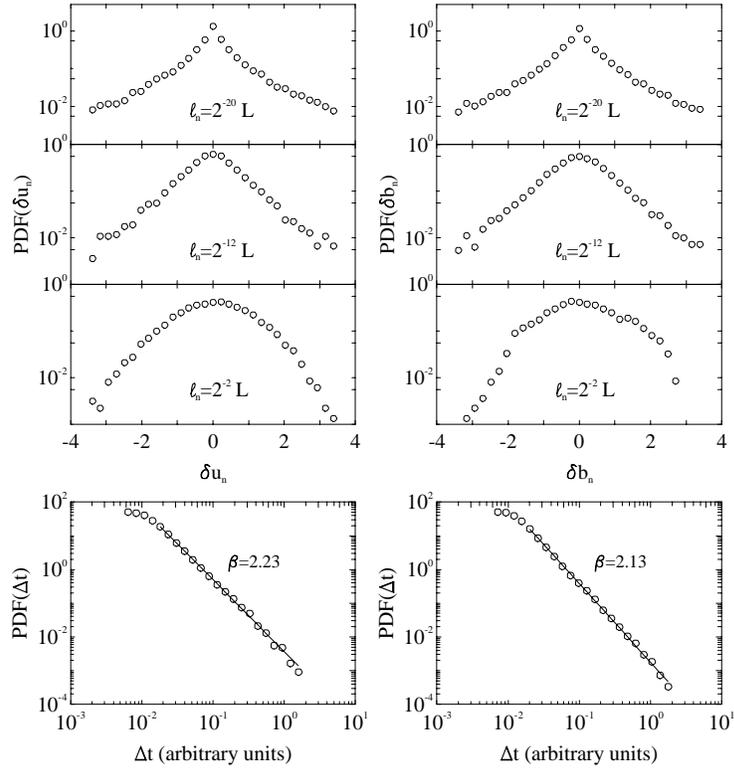}}
\caption{We show results obtained from simulations of the GOY MHD shell model.
Left--hand and right--hand columns refer to velocity and magnetic variables
respectively. In the first three panels of each column we show the pdfs of the
rescaled dynamical variables at three different scales. In the
bottom panels of each column we report the distribution of waiting times
between events at the scale $\ell_n/L = 2^{-20}$ (where $L = k_0^{-1}$). The
values of $\beta$ represent the best fits of the power laws.}
\label{shell}
\end{figure}

The model has a fixed point in the form of a Kolmogorov--like scaling $u_n
\sim b_n \sim k_n^{-1/3}$, and exibits a chaotic dynamics on a strange
attractor in the phase space around the fixed point. This is due to the
presence of the ideal invariants. What is interesting here is the fact that
fluctuations of the degree of chaoticity in the model generate a multifractal
probability measure on the attractor. As a consequence the statistics of a
sample of turbulence coming from simulations of the system is characterised by
multifractal scalings, and this manifest itself as a breakdown of
self--similarity of the system. That is, at variance with sandpile models,
pdfs of the rescaled variables $\delta u_n = Real\{u_n\}/\sigma_{u_n}(\ell_n)$
and $\delta b_n = Real\{b_n\}/\sigma_{b_n}(\ell_n)$ (see figure \ref{shell})
strongly depend on the scale $\ell_n$. At large scales pdfs are gaussian, while
they tend to develop fat tails for small $\ell_n$, towards the dissipative
range. The departure from gaussianity visible at smaller scales, is due to
bursts localised in time, which can be identified and extracted through an
iterative procedure on $|u_n|^2$ and $|b_n|^2$, described extensively in Ref.
\cite{noi}. The WTD for these events is shown in the bottom panels of figure
\ref{shell}. We found a power law $P(\Delta t) \sim \Delta t^{-\beta}$ over at
least two decades (the scaling exponents $\beta$ are reported on the figures).
Long--range correlations between bursts are the origin of the power law for
waiting times.

\begin{figure}
\epsfxsize=12cm
\centerline{\epsffile{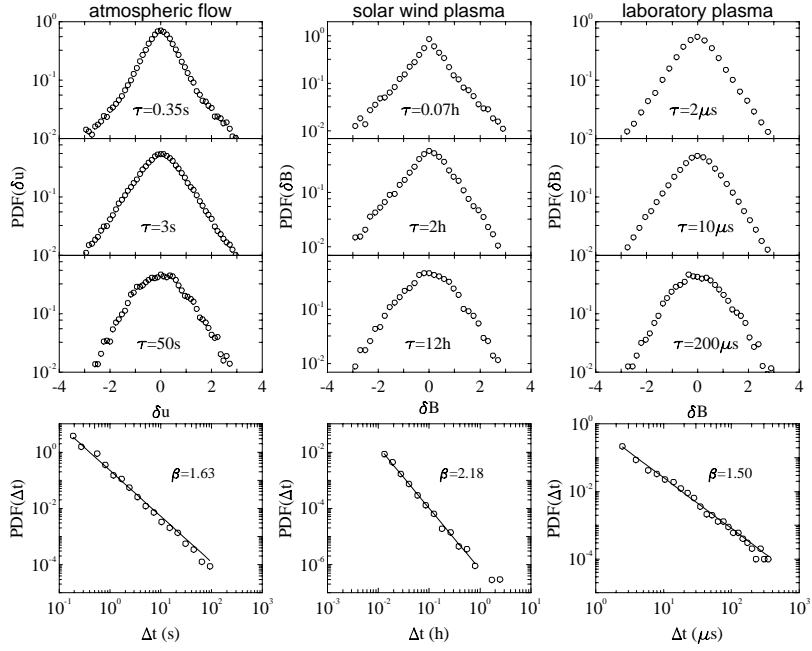}}
\caption{Results obtained from a sample of fluid turbulence (left--hand
column), a sample of magnetic turbulence measured by the Helios II satellite in
the interplanetary space at distance $R = 0.9$ Astronomical Units (central
column), and a sample of magnetic turbulence measured in a magnetically
confined plasma in Reversed Field Pinch configuration (right--hand column). In
the first three panels of each column we show the pdfs of the rescaled
dynamical variables at three different scales. In the bottom panels of each
column we report the distribution of waiting times between events at the
smallest scale. The values of $\beta$ represent the best fits of the power
laws.}
\label{experiments}
\end{figure}

We proceed now to investigate some samples of turbulent flows. First of all we
consider a sample of turbulence as measured in the atmospheric boundary layer
\cite{sample}, where the longitudinal velocity field $u(t)$ is recorded. Then
we consider two samples of magnetic field intensity $B(t)$ in plasmas 
as recorded
both in a given low--speed stream of the solar wind by Helios II spacecraft
\cite{sorriso}, and in a laboratory device for thermonuclear fusion research
(the RFX experiment \cite{rfx}). Even if turbulent samples are collected in
very different physical conditions, they show the same statistical properties.
The rescaled fluctuations, namely $\delta u_{\tau} =
[u(t+\tau)-u(t)]/\sigma_u(\tau)$ and $\delta B_{\tau} =
[B(t+\tau)-B(t)]/\sigma_B(\tau)$, depend on the scale $\tau$ (figure
\ref{experiments}). At large scales pdfs are gaussian, while they tend to
develop fat tails as $\tau \to 0$, towards the dissipative range. The
departure from gaussianity, which results in a correction to the usual
Kolmogorov scaling laws, is due to "intermittency in fully developed
turbulence" (see Ref. \cite{vanatta,frisch} and references therein). Since at
smaller scales the wings of pdfs increase, intermittency is represented by
intense and localised fluctuations which accumulates at smaller scales. These
"events" are generated during the turbulent energy cascade, and can be
interpreted as due to a non homogeneous breakdown of "eddies" towards small
scales \cite{frisch}. Their presence originates the departure from a pure
self--similarity of fluctuations at the various scales. The time of occurrence
of the events in a turbulent flow can be extracted through the procedure
described in Ref. \cite{noi}, and the distribution of waiting times between
them is shown in the bottom panels of figure \ref{experiments} for the
different samples. A power law over two decades, with non universal scaling
exponents $\beta$, is found in all samples.

To summarize, we found that turbulence and the GOY MHD shell model always show
power law for WTD, while this property is shared only by a SOC model, namely
the running sandpile in the ``strong'' feeding regime, far from BTW limit
\cite{soc,Jensen}. Moreover pdfs of rescaled fluctuations in both BTW and
other SOC models do not change shape with the scale $\tau$. On the contrary
intermittency in turbulent flows manifests itself as a breakdown of a pure
self--similarity, leading to increasingly non gaussian tails for pdfs as the
scale becomes smaller. The same phenomenon is observed in the GOY shell model,
and is due to "time intermittency". Non linear terms in the model, which
conserve some few invariants, are responsible for the occurrence of bursts of
chaoticity which are concentrated on the dissipative shells \cite{inter}. Such
typical behaviour in the shell model generates a breaking of the global scaling
invariance of the model, and this results in a correction to the
Kolmogorov--like scaling law. This correction has the same statistical effects
of the corrections of the Kolmogorov scaling laws due to usual intermittency
in real fluid flows. Then as far as the statistical properties we investigated
are concerned, SOC models we examined do not adequately mimic neither fluid nor
plasma turbulence. On the contrary time intermittency in the GOY shell model
allows the model to capture the gross statistical features of real turbulence.

\newpage

\newpage

\newpage

\newpage

\end{document}